# Spatial anisotropy of the exciton level in CaF$_2$ at 11.1 eV and its relation to the weak optical anisotropy at 157 nm

M. Letz,[1] L. Parthier,[2] A. Gottwald,[3] and M. Richter[3]
[1]*Schott Glas, Research and Development, D-55014 Mainz, Germany*
[2]*Schott Lithotec AG, Otto-Schott Strasse 13, D-07745 Jena, Germany*
[3]*Physikalisch-Technische Bundesanstalt, Abbestr. 2-12, D-10587 Berlin, Germany*

CaF$_2$ is the basic optical material for the next step in photolithography to produce nanostructures at a wavelength of 157 nm by the semiconductor industry. Recently an optical anisotropy on CaF$_2$ has been observed at 157 nm which implies serious consequences for the design of the precision optics. In the present work we demonstrate that this optical anisotropy originates from a spatial anisotropy of the exciton level at 11.1 eV as a fundamental effect of the CaF$_2$ crystal with cubic symmetry. For this purpose we have investigated the excitonic state by precision reflection measurements using dispersed synchrotron radiation on oriented surfaces of CaF$_2$ with extremely low impurity concentrations. The results are discussed in terms of the complex dynamic dielectric function.

Fluoride crystals are strongly ionic and have the largest band gaps known for crystalline solids. As a consequence, they are transparent even in the wavelength region of deep ultraviolet (DUV) radiation. Driven by the need of miniaturization from the semiconductor industry, photolithography using DUV radiation is currently applied at a wavelength of 193 nm and under development for 157 nm. For F$_2$ laser radiation at 157 nm, precision transmission optics built from CaF$_2$ are the key components of a lithographic imaging system. For this purpose, pure crystals of CaF$_2$ with extremely low impurity concentrations have been grown at Schott Lithotec. Recently, the effect of an optical anisotropy for CaF$_2$ has been observed at 157 nm which is of great importance for the 157-nm technology.[1] Already the measured weak birefringence of $\Delta n/n \approx 10^{-6}$ of the refractive index considerably influences the design of the imaging optics.

Due to its cubic symmetry, it is generally presumed that the CaF$_2$ crystal shows isotropic optical properties, e.g., for refraction and absorption. However, strongly ionic crystals show deep excitonic bound states. The most pronounced one in CaF$_2$, the $\Gamma$ exciton, is known to arise at 11.1 eV,[2] yielding a strong and narrow absorption structure at a wavelength of about 112 nm. Close to such a strong absorption line, an optical anisotropy can result from a slight deviation of the cubic symmetry, due to the incident radiation field, and a spatial dependence, i.e. the deviation from a spherical $\hbar^2|\mathbf{q}|^2/2m$ dispersion, of the photoabsorption, even when a perfectly cubic crystal is irradiated at optical wavelengths much larger than the lattice constant. The effect of optical anisotropy in the vicinity of a narrow absorption line was formulated by Ginzburg as early as 1958.[3] The relation with exciton physics was pointed out in Ref. 4. The effect is well established at optical wavelengths and usually called "spatial dispersion induced birefringence." It has been measured in the past on various semiconductor materials[5–7] and alkali halide crystals[8] and already explained by exciton anisotropy.

For CaF$_2$, only the optical anisotropy at 157 nm was investigated so far and the relation with exciton dispersion made in theory only.[1,9] In this work we will describe a direct measurement of the spatial anisotropy of the narrow absorption line at 112 nm due to exciton excitation around the $\Gamma$ point. The experiments were performed at the UV and VUV beam line for detector calibration and reflectometry in the Radiometry Laboratory of the Physikalisch-Technische Bundesanstalt, Germany's national metrology institute, at the electron storage ring BESSY (Ref. 10) on extremely pure CaF$_2$ samples from Schott Lithotec. The samples with impurity concentrations well below the ppm level were prepared with surface orientations in the (111) and (100) directions of the crystals. In a reflectometer, the reflectance of the samples was measured in a near-normal incidence geometry. The incoming beam of synchrotron radiation, spectrally dispersed by a normal-incidence monochromator, as well as the reflected beam were measured by the same semiconductor photodiode. The monochromator allows continuous tuning of the wavelength from 40 to 400 nm. In order to enhance the spectral purity of the beam, an Ar gas cell was used to suppress higher diffraction orders from the monochromator grating.

The measurements were performed by embedding the reflection measurements of the samples into two reference measurements of the incoming beam intensity. Online monitoring allowed normalization to variations of the beam intensities. The results of the reflection measurements are shown in Fig. 1. The absolute reflectance values are in good agreement with corresponding results obtained by ellipsometry.[11] When comparing the results of the two different samples, a small but significant shift of about 0.2 nm becomes apparent, the resonance structure of the sample with (111) surface orientation being shifted towards smaller wavelengths. To verify the experimental significance of this small shift, the reproducibility of the wavelength was tested by simultaneously recording resonance absorption lines of Ar in the gas cell during the reference measurements. The Ar absorption can be seen in Fig. 2, where measurements of the Ar resonances at 104.8 and 106.7 nm are shown. The variations of the wavelength scale between the two reference scans before and after the reflection measurement were found to be one order of magnitude smaller than the shift observed in the $CaF_2$ resonance structure.

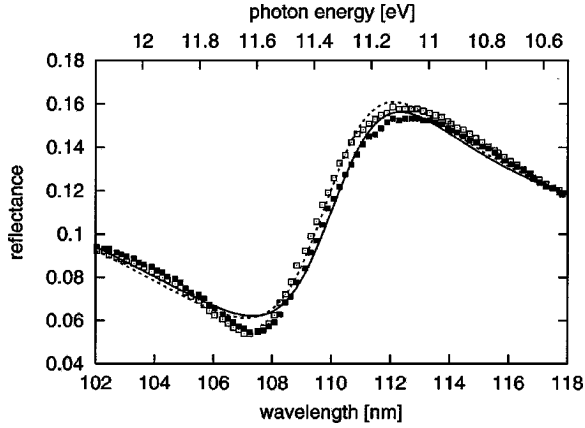

FIG. 1. Normal incidence reflectance around the $\Gamma$ point of $CaF_2$ as a function of wavelength for a (100) surface (experimental data: closed square, fit: solid line) and for a (111) surface (experimental data: open squares, fit: dashed line).

For interpreting our experimental data with regard to exciton position and lifetime, we chose the following expression to connect the reflectance with the complex dynamic dielectric function:

$$R(\omega) = \frac{|\tilde{n}(\omega) - 1|^2}{|\tilde{n}(\omega) + 1|^2}, \qquad (1)$$

where the complex dynamic refractive index $\tilde{n} = n + ik$ is related to the dynamic dielectric function by $\tilde{n}(\omega)^2 = \epsilon(\omega)$. The complex dielectric function was approximated by

$$\epsilon(\omega) = \epsilon_\infty + \frac{a}{\omega_0^2 - \omega^2 - i\omega/\tau} + \epsilon_{band}(\omega), \qquad (2)$$

where $\epsilon_\infty$ denotes the basically constant, real contribution of the polarizability in the band gap, $a$ is the strength of the excitonic resonance, $\omega_0$ and $\tau$ are exciton position and lifetime, respectively, and $\epsilon_{band}(\omega)$ takes care of the fact that the resonance occurs at a position where the reflectance is already steeply increasing due to the vicinity of trans bandgap excitations. The approximation follows Maxwell's equations with a semi-infinite excitonic wave function in the bulk system, neglecting any surface effects such as a dead zone of vanishing excitonic wave functions[12] or the surface roughness of the samples. Moreover, the shape of our reflection signal does not suggest the need for additional boundary conditions[13] within the framework of a more sophisticated treatment.[14,15]

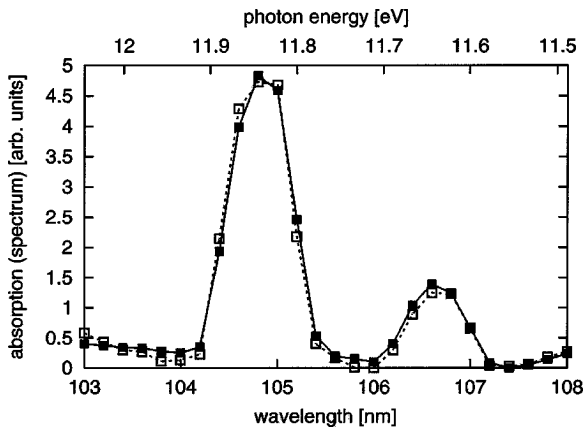

FIG. 2. Photoabsorption spectrum of the two Ar resonance lines at 104.8 and 106.7 nm as observed in two independent reference measurements shown by the closed and open symbols, respectively.

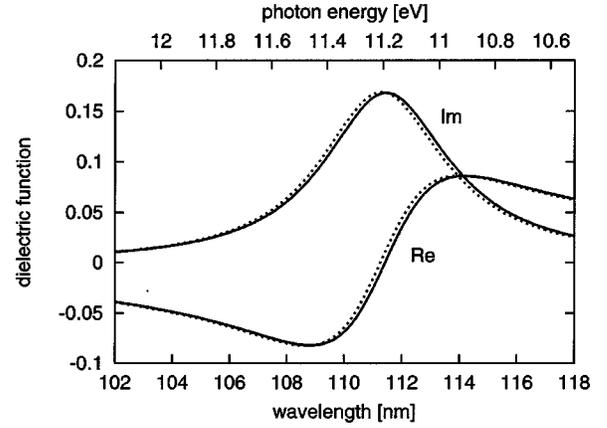

FIG. 3. Real (Re) and imaginary (Im) part of the contribution to the dielectric function resulting from excitonic resonance [second part in Eq. (2)] normalized to equal excitonic resonance strength as derived from fits to the experimental data in Fig. 1 (solid line: 100 orientation, dashed line: 111 orientation).

According to Eqs. (1) and (2), we applied a least squares fitting algorithm to determine the parameters $a$, $\omega_0$, and $\tau$. The value of $\epsilon_\infty = n_\infty^2 = 1.4^2$ was taken from the known real refractive index $n_\infty$ in the band gap. Furthermore, the band contribution $\epsilon_{band}(\omega)$ was adjusted. The fitting procedure allows an accurate determination of exciton position and lifetime. The resulting dielectric functions are shown in Fig. 3. They are normalized to equal excitonic resonance strength $a$ because in the present work exciton position and lifetime are the only parameters of interest. As a result one obtains for the exciton position of the different crystal orientations

$$\hbar\omega_0^{(100)} = 11.111 \text{ eV}, \quad \lambda_0^{(100)} = 111.67 \text{ nm},$$
$$\hbar\omega_0^{(111)} = 11.131 \text{ eV}, \quad \lambda_0^{(111)} = 111.47 \text{ nm}. \qquad (3)$$

The lifetime at room temperature results at $\tau \approx 5$ fs and the magnitude of the resonance shift between the two spatial orientations at

$$\Delta\hbar\omega_0 = (0.020 \pm 0.002) \text{ eV}, \quad \Delta\lambda_0 = (0.20 \pm 0.02) \text{ nm}. \qquad (4)$$

We will now discuss the measured shift of the excitonic resonance in terms of the recently measured weak optical anisotropy in $CaF_2$.[1] In this context, we apply an expansion of the dielectric function around zero momentum. Due to the divergence of the dielectric function, this expansion will not yield a correct expression in the vicinity of the exciton resonance, but it allows a consistency check for both anisotropy effects. Expanding the dielectric tensor up to the second order in wave vector **q** for a system with inversion symmetry where the term linear in **q** vanishes results in

$$\epsilon_{ij}(\omega) = \epsilon_0(\omega)\delta_{ij} + \alpha_{ijkl}(\omega)q_k q_l + 0q^4. \quad (5)$$

The fourth rank tensor $\alpha_{ijkl}(\omega)$ has to be symmetric when interchanging the first two or last two indices, respectively, $\alpha_{ijkl} = \alpha_{jikl} = \alpha_{ijlk}$.[16] Further restrictions are due to the cubic symmetry of the crystal: $\alpha_{1111} = \alpha_{2222} = \alpha_{3333}$, $\alpha_{1122} = \alpha_{1133} = \alpha_{2211} = \alpha_{2233} = \alpha_{3311} = \alpha_{3322}$, and $\alpha_{1212} = \alpha_{2121} = \alpha_{1313} = \alpha_{3131} = \alpha_{2323} = \alpha_{3232}$. Degrees of freedom are further reduced, taking into account that the wave vector **q** always points into the direction of the perturbation, which yields the additional restriction $\alpha_{1111}:\alpha_{1122}:\alpha_{1212} = 2:-1:-1$.[17] In the following, we will discuss the real part of the dielectric function $\epsilon'_{ij}(\omega)$ and small differences of it in different spatial directions for energies well below the excitonic excitation, where the contribution of the imaginary part is small. We therefore consider the real parts of the tensor $\alpha_{ijkl}(\omega)$ only. For the geometry of the present reflection measurements, the effect of birefringence vanishes for radiation with wave vectors in the (100) and (111) crystal orientations. However, the refractive index along these directions, independent from the polarization direction of the radiation, results in

$$n_{(100)} \approx \sqrt{\epsilon_0 + \alpha_{1122}q^2},$$
$$n_{(111)} \approx \sqrt{\epsilon_0 + (\alpha_{1111} + 2\alpha_{1122} - 2\alpha_{1212})q^2/3}, \quad (6)$$

which follows from diagonalizing the dielectric function for the two directions (100) and (111) of the wave vector **q**. Introducing $\alpha$, which is defined by $\alpha_{1111} = 2\alpha$, and hence $\alpha_{1122} = \alpha_{1212} = -\alpha$, yields

$$n_{(100)} \approx \sqrt{\epsilon_0 - \alpha q^2}, \quad n_{(111)} \approx \sqrt{\epsilon_0 + \frac{2}{3}\alpha q^2}. \quad (7)$$

The indices in brackets still denote the direction of the wave vector of the radiation. We have measured the excitonic resonance occurring for radiation reflected at the (111) surface at a photon energy 0.02 eV above (corresponding to a wavelength 0.2 nm below) the resonance along the (100) surface. This implies, for photon energies below the excitonic resonance, a smaller real part of the dielectric function and refractive index for radiation propagating in the (111) direction of the crystal compared to the (100) direction ($n_{(111)} < n_{(100)}$) and Eq. (7) results in a negative sign of $\alpha$.

On the other hand, for a birefringence measurement with radiation propagating in the (110) direction of the crystal,[1] the radiation with a polarization in the 001 and 1̄10 directions has a refractive index given by

$$n_{001} \approx \sqrt{\epsilon_0 + \alpha_{1122}q^2},$$
$$n_{1̄10} \approx \sqrt{\epsilon_0(\alpha_{1111} + \alpha_{1122} - 2\alpha_{1212})q^2/2}. \quad (8)$$

Here, the indices of the refractive index denote the polarization direction of the radiation. Introducing the parameter $\alpha$ as defined above leads to

$$n_{001} \approx \sqrt{\epsilon_0 - \alpha q^2}, \quad n_{1̄10} \approx \sqrt{\epsilon_0 + \frac{3}{2}\alpha q^2}. \quad (9)$$

The negative sign of $n_{1̄10} - n_{001}$ as measured by Burnett et al.[1] again implies a negative sign for $\alpha$. Thus, the direction of the shift for the excitonic resonance as measured in the present work is consistent with the sign of the optical anisotropy as measured by Burnett et al. at 157 nm.

In conclusion, we have measured the spatial anisotropy of the exciton level around the $\Gamma$ point in CaF$_2$ at 11.1 eV on oriented samples of high purity. The results are discussed in terms of the complex dynamic dielectric function and explain the optical anisotropy at 157 nm by the interaction of radiation with the cubic structure of a perfect CaF$_2$ crystal free from impurities or distortions.

M. L. acknowledges helpful discussions with J. H. Burnett, E. G. Tsitsishvili, and C. Klingshirn. The work was supported by the BMBF project "Laserbasierte Ultrapräzisionstechnik–157-nm-Lithographie, Teilvorhaben: Optische Materialien und Komponenten für die 157-nm-Lithographie, AP 4210."